# From Spectroscopy to the Strong Coupling Constant with Heavy Wilson Quarks

Matthew Wingate, Thomas DeGrand

*Department of Physics, University of Colorado, Boulder, CO 80309, USA*

Sara Collins, Urs M. Heller

*Supercomputer Computations Research Institute, Florida State University, Tallahassee, FL 32306-4052, USA*

(January 26, 1995)

## Abstract

In this work we present lattice calculations of the masses of P-wave mesons using Monte Carlo simulations. Our valence fermions are defined by the Wilson action. Our gauge fields are generated with both dynamical staggered fermions at a lattice coupling $\beta \equiv 6/g^2 = 5.6$ for sea quark masses of $am_q = 0.010$ and $0.025$, and in the quenched approximation at $\beta = 6.0$. We present results for charm and charmonium spectroscopy and use them to compute the strong coupling constant $\alpha_s$. We compare our results to those of other recent lattice calculations and experiments.

1

# I. INTRODUCTION

This paper presents results of P-wave meson masses using Wilson quarks on both quenched gauge configurations and configurations which include light dynamical quarks. Here we focus on systems containing one or two heavy quarks.

Lattice calculations of P-wave systems go back only a few years. Simulations with staggered fermions regularly measure the masses of states which are odd-parity partners of the ordinary ground state mesons. The earliest calculations with Wilson fermions of which we are aware were done by the APE collaboration [1] and then in 1992 one of us [2] presented incomplete calculations of the whole P-wave multiplet, which hinted at the existence of fine structure splitting in charmonium in qualitative agreement with experiment. Also in 1992 the Fermilab group performed P-wave spectroscopy for heavy quark systems with Wilson fermions and with improved Wilson fermions [3]. Since then very precise calculations in heavy quark systems have been done using simulations with nonrelativistic quarks by [4–6], with much higher accuracy than Wilson fermions permit, for the same amount of computer time. Indeed, the calculations we present do not show statistically significant fine structure splitting at our heaviest quark masses.

Why are more calculations with Wilson fermions being presented? There are several reasons. First of all, the spectroscopy of states with light quarks demands a lattice implementation of relativistic quarks, so in $D$ meson spectroscopy (for example) at least one of the quarks must be relativistic. The heavy quark may be treated on the lattice as static, or nonrelativistic, or relativistic but heavy. Each prescription has its own set of systematic errors, and so it is important to do the calculation in all three ways on the lattice to understand them. The calculations done in this paper for charm spectroscopy complement ones using the same gauge configurations in a companion work [6] using nonrelativistic heavy quarks.

Second, even without particularly accurate P-wave mass measurements we can address a topical physics problem: the extraction of the strong coupling constant from lattice QCD.

The Fermilab group [3] was the first to use the S-P mass splitting of heavy mesons to set the scale, which allows one to run the coupling calculated from the plaquette to any scale. However, since their calculation was in the quenched approximation, they had to correct for the presence of dynamical fermions. Our simulations are done in the presence of dynamical staggered fermions, so we avoid this extrapolation. Similar calculations have been done in two-flavor QCD [7] and using non-relativistic quarks [4]. Further details regarding the simulations are given in Section II.

We use interpolating fields which are generalizations of local currents to couple to the two S-wave states and to one of the P-wave states (the $^3P_1$). However, in order to create all of the P-wave states, we employ several representations of the lattice cubic group. Both of these procedures are described in Section III.

In Section IV we describe in detail our methods for extracting the strong coupling from measurements of the plaquette, for using zero flavor and two flavor calculations of $\alpha_s$ to extrapolate to three flavors, for changing from our lattice definition of $\alpha_s$ to the modified minimal subtraction definition, and for perturbatively running the coupling to any momentum scale. Particular attention is given to possible sources of systematic errors in our calculation.

Finally, in Section V we present lattice calculations of the masses of charmed mesons. Our computations confirm that the Wilson action does not generate the hyperfine structure correctly.

## II. THE SIMULATIONS

Our dynamical fermion simulations were carried out on the Connection Machine CM-2 at the Supercomputing Computations Research Institute at Florida State University, and our quenched simulations on the Paragons at the San Diego Supercomputer Center and at Indiana University.

For the dynamical simulation we used the ensemble of configurations generated by the HEMCGC collaboration with two flavors of dynamical staggered quarks [8]. The configurations were generated using the hybrid molecular dynamics (HMD) algorithm [9]. The size of the lattices is $16^3 \times 32$, the lattice coupling is $\beta = 5.6$, and the dynamical quark masses are $am_q = 0.010$ and $0.025$. Periodic boundary conditions were used in all four directions of the lattice. The total simulation length was 2000 simulation time units (with the normalization of ref. [10]) at each quark mass value. We analyzed lattices spaced by 20 HMD time units, for a total of 100 lattices at each mass value.

The spectroscopy was computed with six values of the Wilson quark hopping parameter: $\kappa = 0.1600, 0.1585, 0.1565, 0.1525, 0.1410$, and $0.1320$. The first three values are rather light quarks (the pseudoscalar mass in lattice units ranges from about 0.25 to 0.45), and the other three values correspond to heavy quarks (pseudoscalar mass from 0.65 to 1.5). Our inversion technique is conjugate gradient with preconditioning via incomplete lower-upper (ILU) decomposition by checkerboards [11]. For more details about the dynamical staggered fermion simulations see ref. [8,12]. Since we use sources for the propagators which are extended in space, we fix gauge to lattice Coulomb gauge using an overrelaxation method [13].

The quenched configurations were generated using an updating algorithm which treats each link to a combination of four microcanonical overrelaxed hits on each of three SU(2) subgroups, followed by one Kennedy-Pendleton [14] quasi-heat bath update. The lattices are $16^3 \times 48$ at a lattice coupling $\beta = 6.0$. Lattices were recorded for analysis every 200 sweeps, and we acquired a total of 79 lattices. Here, the spectroscopy was computed with five values of the Wilson hopping parameter: $\kappa = 0.1550, 0.1540, 0.1520, 0.1450$, and $0.1300$. Our boundary conditions, gauge fixing, and inversion technique are identical to those mentioned above.

## III. P-WAVE SPECTROSCOPY

We extract masses from correlation functions of some operator $O$,

$$C(t) = \langle 0|O(t)O(0)|0\rangle. \tag{1}$$

which for large $t$ reduces to a single decaying exponential

$$C(t) \simeq \left|\langle 0|O|1\rangle\right|^2 e^{-amt}, \qquad (2)$$

where $m$ is the mass of the lightest particle that couples to $O$. We employ two types of operators to couple to the P-wave states: operators whose spatial dependence is symmetric and which couple to the desired state through the quarks' spins ("current" operators), and operators which form specific representations of the cubic group.

## A. "Current" Operators

The first class of operators we use are the analogs of currents. In this work we use the pseudoscalar, vector, and axial-vector currents given respectively by

$$P = \bar{\psi}\gamma_5\psi, \qquad (3)$$

$$V_i = \bar{\psi}\gamma_i\psi, \qquad (4)$$

$$A_i = \bar{\psi}\gamma_i\gamma_5\psi. \qquad (5)$$

These currents have respective quantum numbers $J^{PC}$ as follows: $0^{-+}$, $1^{--}$, and $1^{++}$. In spectroscopic notation, $^{2S+1}L_J$, which we shall use hereafter, these states are the $^1S_0$, $^3S_1$, and $^3P_1$, respectively.

For the present work we consider correlators which have "shell" sources and "point" sinks. That is, we measure the source of the correlator over a spatial volume with a Gaussian distribution centered at a point, and the sink at a single point. We also have data with "shell" sinks, from which the masses are consistent with, but noisier than, the point sink data. In either case the sink state is projected onto $\vec{k} = 0$.

One can examine the quality of the signal by calculating "effective" masses: fits to the correlator over just two neighboring points. If the assumption that the current is coupling to just one particle state is correct, then we should see a plateau in the effective mass. Once we establish an approximate range over which to fit, we fit every possible range therein and select the best one, in our case the one which maximizes the confidence level times the number of degrees of freedom divided by the statistical error of the mass. (This selection criterion has been used in our previous studies [8,12].)

Effective mass plots and best fits to the dynamical staggered fermion data for the S-wave states have been published in a previous work [8]. In Figures 1 and 2 we show effective masses for the $^3P_1$ state for $am_q = 0.010$ and $0.025$. Note that, unlike the S-wave states, we can only fit out to distances $t \sim 10$ since we lose the signal in noise. The best range fits are listed in Tables I-V and VI-X.

Results from the quenched simulations are presented here as follows: effective mass plots for the pseudoscalar, vector, and axial-vector currents are shown in Figures 3-5, and our best range fits are presented in Tables XI-XIII.

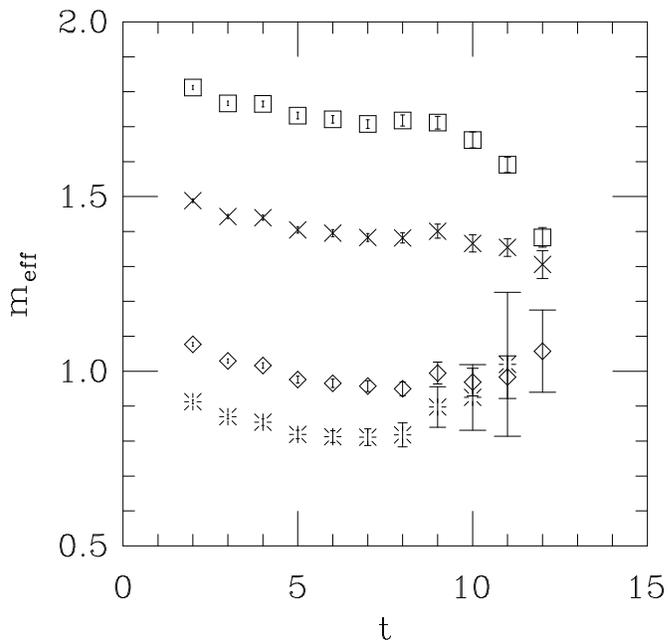

FIG. 1. Effective masses for the $^3P_1$ state using $A_i = \bar{\psi}\gamma_i\gamma_5\psi$ for $am_q = 0.010$ at $\kappa_q = \kappa_{\bar{q}} = 0.1320, 0.1410, 0.1525, 0.1565$.

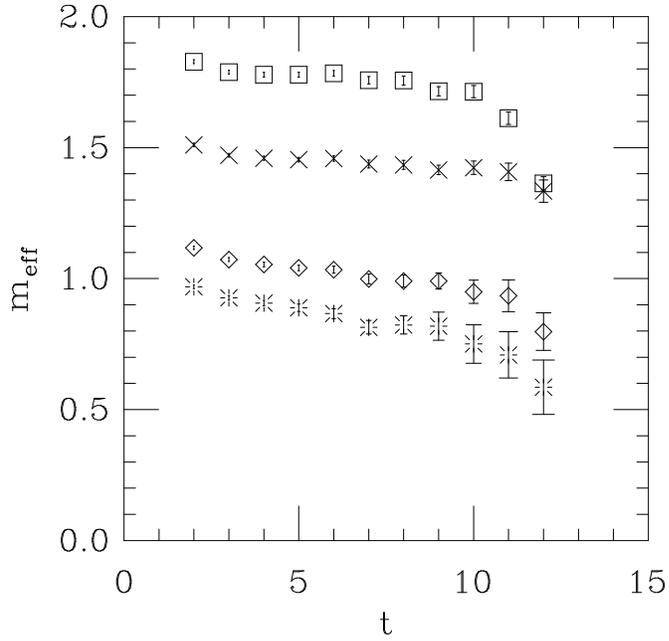

FIG. 2. Effective masses for the $^3P_1$ state using $A_i = \bar{\psi}\gamma_i\gamma_5\psi$ for $am_q = 0.025$ at $\kappa_q = \kappa_{\bar{q}} = 0.1320, 0.1410, 0.1525, 0.1565$.

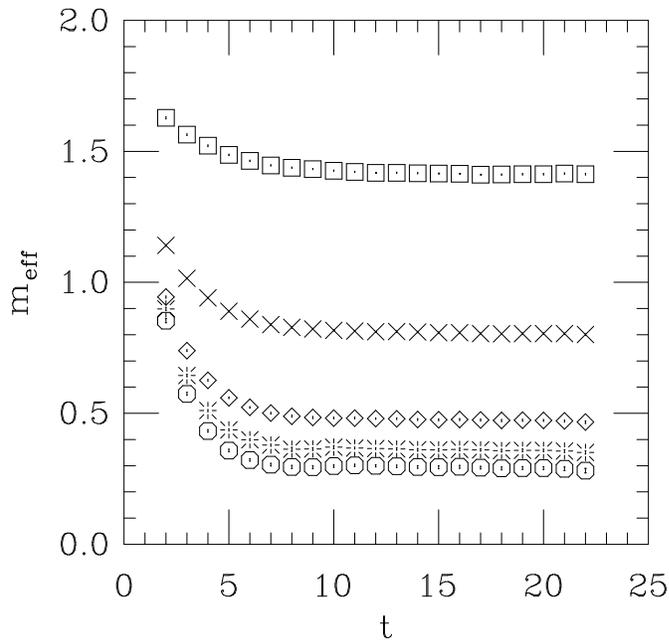

FIG. 3. Effective masses for the $^1S_0$ state using $P = \bar{\psi}\gamma_5\psi$ for the quenched data at $\kappa_q = \kappa_{\bar{q}} = 0.1300, 0.1450, 0.1520, 0.1540, 0.1550$.

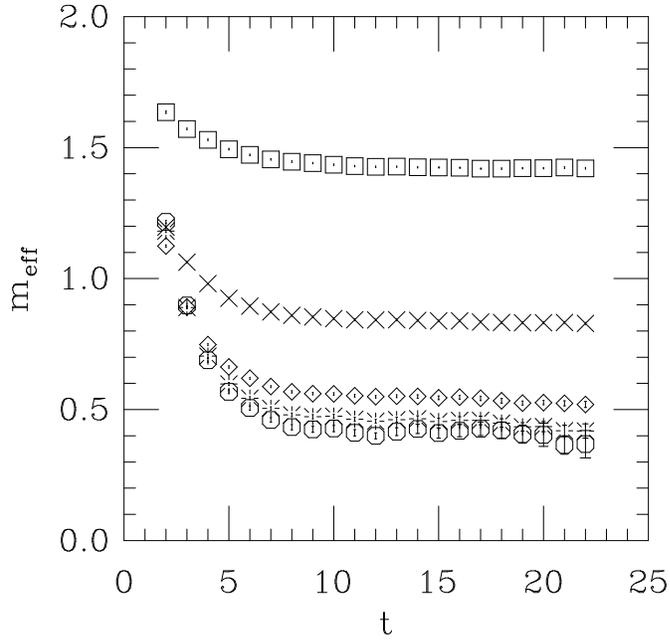

FIG. 4. Effective masses for the $^3S_1$ state using $V_i = \bar{\psi}\gamma_i\psi$ for the quenched data at $\kappa_q = \kappa_{\bar{q}} = 0.1300, 0.1450, 0.1520, 0.1540, 0.1550$.

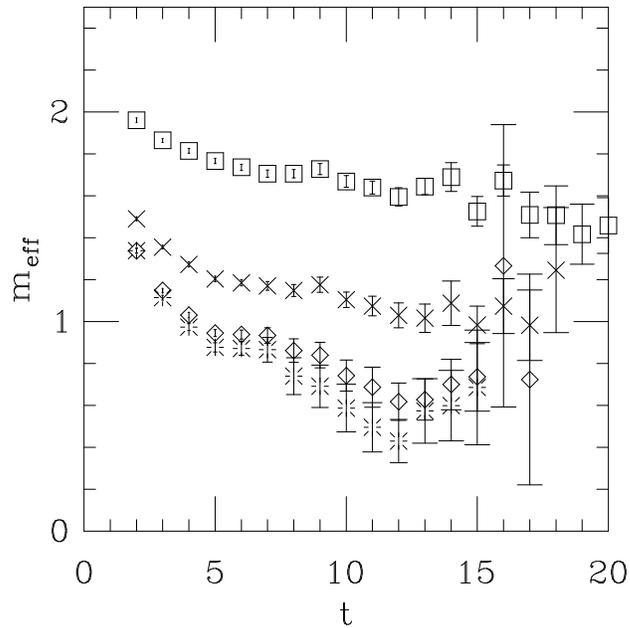

FIG. 5. Effective masses for the $^3P_1$ state using $A_i = \bar{\psi}\gamma_i\gamma_5\psi$ for the quenched data at $\kappa_q = \kappa_{\bar{q}} = 0.1300, 0.1450, 0.1520, 0.1540$. The signal for the $\kappa = 0.1550$ state is noisy and not shown here.

## B. Cubic Group Representations

On the lattice, instead of the angular momentum group, the cubic group is the symmetry group according to which states form irreducible representations. The "current" operators previously discussed have "orbital cubic group" representation $A_1$, corresponding to "orbital angular momentum" representation $L = 0$. The "spin cubic group" representations are $A_1$ for the pseudoscalar and $T_1$ for vector and axial vector currents.

We can also build the equivalent of "orbital angular momentum" $L = 1$, *i.e.*, P-wave, states on the lattice. They are in the cubic group representation $T_1$ and have wave functions $xf(r)$, $yf(r)$ and $zf(r)$ with $f(r)$ a function depending only on the radius. In our case $f(r)$ is taken as a Gaussian. Equivalently, the wave functions can be taken as $(x+iy)f(r)$, $(x-iy)f(r)$ and $zf(r)$. On a finite lattice, in order to satisfy periodic boundary conditions in the spatial directions, we use $\sin(2\pi x_k/L_k)$ instead of $x_k$ in the wave functions.

This "orbital cubic group" representation can be combined with the two different "spin cubic group" representations to get

$$T_1 \otimes A_1 = T_1, \tag{6}$$

$$T_1 \otimes T_1 = A_1 + T_1 + E + T_2 \tag{7}$$

The first line corresponds to a continuum $^1P_1$ state and the second to continuum $^3P_0$, $^3P_1$ and $^3P_2$ states. The lattice $E$ and $T_2$ states combine in the continuum limit to become the $^3P_2$ states.

In the numerical measurements we use one "orbital P-wave" (cubic group $T_1$) source, the $x+iy$ one. We use all spin combinations at the source and sink and all "orbital cubic group $T_1$" wave functions at the sink. With this we can build 7 of the 9 possible correlation functions of the $^3P$ states. One each of the three $T_1$ and $T_2$ states are inaccessible with our single $x+iy$ source. In addition one can insert extra $\gamma_0$'s at either sink and/or source. We implemented only the variant with $\gamma_0$'s at both source and sink. However, the result turned out to be somewhat noisier than without these $\gamma_0$ factors in the operators and thus we will ignore them in the rest of the paper.

Again, we compare effective mass fits and fits to a range of points. In this case we have several operators for each combination of spin and parity. Generally one or two operators provide a superior signal compared to the other ones.

We will restrict our presentation of tabular masses to the following states: for the $^3P_2$ state we use one of the two $E$ states, which we will refer to as $E_a$: $\frac{1}{\sqrt{2}}(O_{++} + O_{--})$. Here the first subscript refers to the "spin" content with $\pm$ standing for $\gamma_\pm$ and 0 for $\gamma_3$, while the second subscript refers to the orbital angular momentum content with $\pm$ standing for $x \pm iy$ and 0 for $z$, up to normalization factors. For $^3P_1$ we use the $T_{1a}$ state, $\frac{1}{\sqrt{2}}(O_{0+} - O_{+0})$, for $^3P_0$ the $A_1$ state, $\frac{1}{\sqrt{3}}(O_{+-} + O_{-+} + O_{00})$, and for the $^1P_1$ we use the cubic group $^1P_1$ state, $O_{5+}$, where the 5 refers, of course, to $\gamma_5$. Note, again, that all the states used couple to the orbital $x + iy$ source. The lattice masses for best fits for each different spin-parity combination are listed in Tables II-V for the $am_q = 0.010$ data and in Tables VII-X for the $am_q = 0.025$ data.

We found that the different representations of the cubic group corresponding to the same angular momentum states are degenerate: the lattice does not break rotational symmetry so badly that we can observe it in spectroscopy. Fig. 6 illustrates this point: it is a correlated fit to the mass difference of the $E_a$ and $T_{2a}$ ($\frac{1}{\sqrt{2}}(O_{0+} + O_{+0})$) states, both of which correspond to $^3P_2$ continuum states.

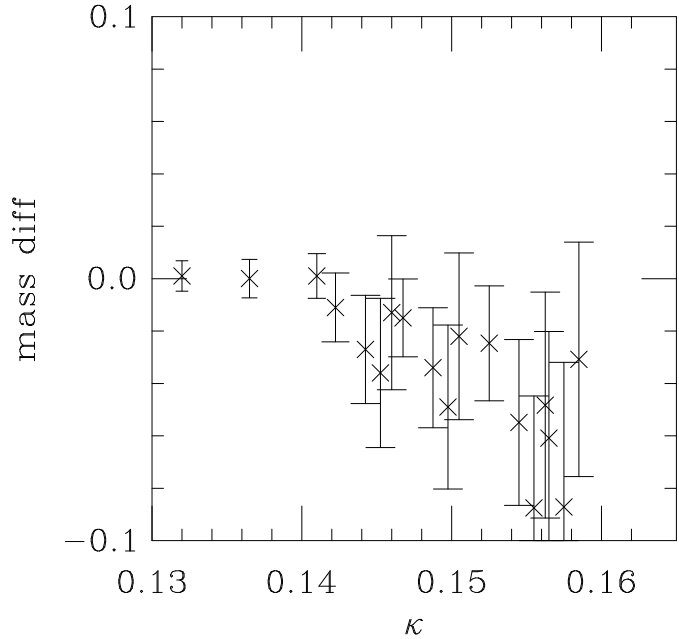

FIG. 6. Mass difference between two different representations of the cubic group corresponding to the $^3P_2$ meson, from the $am_q = 0.010$ data set as a function of the average quark hopping parameter at $\kappa = (\kappa_q + \kappa_{\bar{q}})/2$.

All lattice data are correlated, and so to look for fine structure by simply subtracting the two masses overestimates uncertainties. Instead, we perform a correlated fit of the two propagators to two masses and extract an uncertainty from the error matrix. We present pictures of these differences in Figs. 7-9 (for the $^3P_2 - {}^3P_1$, $^3P_1 - {}^3P_0$, and $^3P_1 - {}^1P_1$ mass differences, respectively). For the heaviest quark masses we are unable to see any fine structure splitting within our statistical uncertainty.

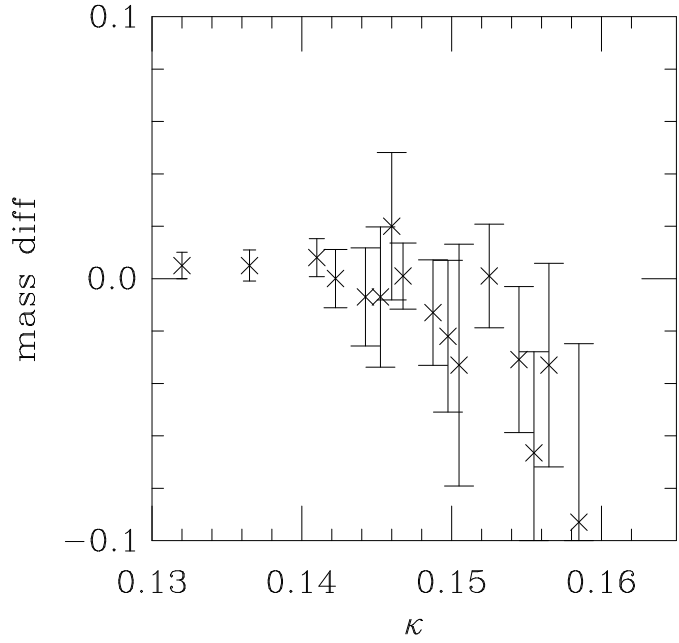

FIG. 7. $^3P_2 - {}^3P_1$ meson mass difference from the $am_q = 0.01$ data set as a function of the average quark hopping parameter at $\kappa = (\kappa_q + \kappa_{\bar{q}})/2$.

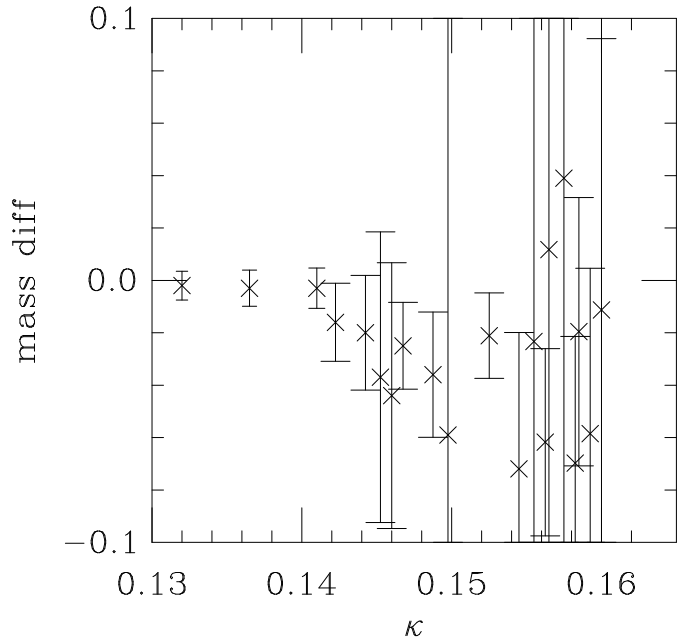

FIG. 8. $^3P_1 - {}^3P_0$ meson mass difference from the $am_q = 0.01$ data set as a function of the average quark hopping parameter at $\kappa = (\kappa_q + \kappa_{\bar{q}})/2$.

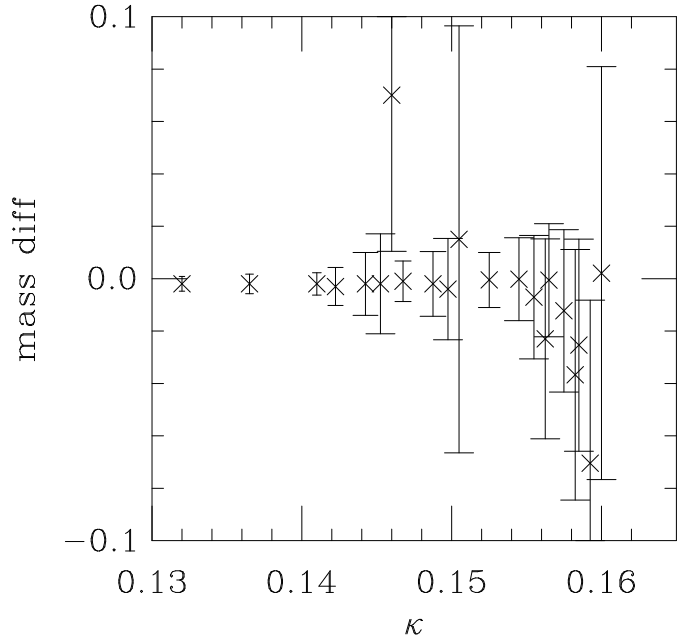

FIG. 9. $^3P_1 - {}^1P_1$ meson mass difference from the $am_q = 0.01$ data set as a function of the average quark hopping parameter at $\kappa = (\kappa_q + \kappa_{\bar q})/2$.

## IV. THE STRONG COUPLING CONSTANT

The lattice allows a determination of the strong coupling constant through the measurement of two physical and nonperturbative quantities. The mass splitting between two states, computed on the lattice, gives a length scale $a$, the lattice spacing. The plaquette (or any other short distance quantity) gives a coupling constant at a momentum scale inversely proportional to the lattice spacing.

The method uses a definition of the strong coupling constant in terms of a physical observable [15]: that is, we define $\alpha_V(q)$ through the potential

$$V(q) \equiv -\frac{C_f 4\pi \alpha_V(q)}{q^2}. \tag{8}$$

Here $C_f = 4/3$ is a group theory factor, and $q$ is the gluon momentum. With this definition, the perturbative expression for the logarithm of the trace of the plaquette is [12,15]:

$$-\log\left\langle \frac{1}{3}\text{Tr}\, U_{plaq} \right\rangle = \frac{4\pi}{3}\alpha_V(3.41/a)\left[1 - (1.19 + 0.07 n_f)\alpha_V + \mathcal{O}(\alpha_V^2)\right]. \tag{9}$$

Once the lattice spacing and the coupling constant are fixed one may convert the coupling constant to other schemes and run it to other desired scales.

Of course, this calculation is incomplete. It depends on the empirical observation that the coupling constant defined through the potential provides a good perturbative expansion parameter from fairly large lattice spacing on down. A completely nonperturbative alternative is the calculation of the strong coupling constant by Lüscher, Weisz, Wolff, and collaborators [16]. Here the coupling constant is defined via the response of a lattice system to an external color electric field, and a wide range of physical scales can be covered by a series of steps each of which involves only a small change of physical scale. From a phenomenological point of view the two methods produce equivalent results for $\alpha_{\overline{MS}}(M_Z)$ from quenched simulations. This program has not been carried out for full QCD yet [17].

## A. Fixing the Lattice Spacing

To set the scale we use the mass difference between P-wave states ($L = 1$) and S-wave states ($L = 0$) in the charmonium sector. We choose this physical quantity, rather than the mass of the rho meson for example, for two reasons. First, the S-P splitting is fairly insensitive to the quark masses. Second, in the laboratory the widths of the charmonium states are much narrower than those of the light mesons. Therefore, it is realistic to think that the S-P mass splitting is less sensitive to lattice artifacts than are light hadron masses.

Using currents to couple to the pseudoscalar, vector, and axial vector mesons, we can compute the mass difference

$$\Delta M(^3P_1 - \overline{S}) \equiv M(^3P_1) - \frac{1}{4}\Big[M(^1S_0) + 3M(^3S_1)\Big] \tag{10}$$

We perform a six parameter correlated fit to the three propagators in order to calculate this mass difference [18] using a fit function for each propagator of

$$f_i(t) = A_i\Big(e^{am_i t} + e^{am_i(N_t - t)}\Big). \tag{11}$$

Since we are going to fix our measured lattice splitting to the mass splitting in charmonium, we need to have a rough idea of $\kappa_{\text{charm}}$. In previous work on these configurations [8], the lattice spacing was determined by fixing the rho to its physical value, yielding $1/a = 2140$ and 2000 MeV for sea quark masses 0.010 and 0.025 respectively. To find the approximate value for $\kappa_{\text{charm}}$ we use these lattice spacings to find the approximate $\kappa$ for which the $J/\psi$ mass is at its physical value, i.e. $M_\psi a \simeq 3.1/2.0 = 1.55$. Also, since the charm quark mass is of the same order as the inverse lattice spacing ($m_c \not\ll a^{-1}$), the dispersion relation for Wilson fermions is distorted:

$$E(\vec{k}) = m_1 + \frac{\vec{k}^2}{2m_2} + \ldots \tag{12}$$

with $m_1 \neq m_2$. To account for this we must shift our lattice mass [19,20], $\mu$, using

$$M_\psi a = \mu - 2m_1 a + 2m_2 a, \tag{13}$$

where, with $\tilde{\kappa} = \kappa/(8\kappa_{\text{crit}})$,

$$m_1 a = \log\left(\frac{1 - 6\tilde{\kappa}}{2\tilde{\kappa}}\right), \tag{14}$$

and

$$m_2 a = \frac{\exp(m_1 a) \sinh(m_1 a)}{1 + \sinh(m_1 a)}. \tag{15}$$

$\kappa_{\text{crit}}$ is estimated from [8] to be 0.161 for both values of sea quarks, and 0.157 for the quenched data. This gives us the estimate that for the dynamical fermion data set $\kappa_{\text{charm}}$ is between 0.1320 and 0.1410, the values of our two heaviest flavors of valence Wilson quarks, and just below 0.1300 for the quenched data set. Of course, this was done with a rough guess for the lattice spacing, which is indeed the quantity we are attempting to calculate. However, since we argue that the S-P splittings are insensitive to the quark mass in the heavy systems, we need only a rough idea of the value of $\kappa_{\text{charm}}$.

Therefore, for a given pair of $\kappa$ values, we fit the correlators using pseudoscalar, vector, and axial vector currents to (11). When calculating the mass difference we fit the three propagators together in order to compensate for correlations due to using the same configurations for all three quantities. The results for the quarkonium ($\kappa_q = \kappa_{\bar{q}}$) pairs are listed in Table XIV for dynamical staggered fermion mass = 0.010, Table XV for fermion mass = 0.025, and in Table XVII for the quenched approximation.

Since we cannot know *a priori* the value of $\kappa_{\text{charm}}$, and since we assume $\Delta M$ to be insensitive to $\kappa$ for heavy mesons, we average $\Delta M$ for the two heaviest quarkonium pairs and set them equal to the physical value $M\chi - (M_\psi + 3M_{\eta_c})/4 = 442$ MeV [21] to find the lattice spacing. Again, one should refer to Tables XIV, XV, and XVII for results.

Instead of using an interpolating field to couple to the $^3P_1$ state, we can couple to the full P-wave system by using the corresponding lattice multiplet. Thus, we can also compute the mass differences

$$\Delta M(^1P_1 - \overline{S}) \equiv M(^1P_1) - \frac{1}{4}\Big[M(^1S_0) + 3M(^3S_1)\Big], \qquad (16)$$

and the true "spin-averaged S-P mass splitting",

$$\Delta M(\overline{P} - \overline{S}) \equiv \frac{1}{12}\Big[5M(^3P_2) + 3M(^3P_1) + M(^3P_0) + 3M(^1P_1)\Big] - \frac{1}{4}\Big[M(^1S_0) + 3M(^3S_1)\Big] \qquad (17)$$

To do this, we take the best operator corresponding to each $J^{PC}$ value and in the case of (16) perform a three-propagator correlated fit, or in the case of (17) a jackknife fit. So far this has been done for the dynamical fermion simulations only. The lattice mass differences are presented in Tables XIV and XV.

### B. Lattice Spacings

Tables XIV and XV give a summary of the inverse lattice spacings obtained from the S-P mass splittings. Each mass splitting gives a lattice spacing, thus we must decide which value to use to set the scale. For the $am_q = 0.010$ data set, we note two things: the lattice values for the $P - S$ difference are the same which implies that the $^3P_1$ and $^1P_1$ states are degenerate using Wilson quarks. However this is not the case in the physical world; *i.e.* using the $^3P_1$ state to fix the scale rather than the $^1P_1$ state changes our determination of the lattice spacing. Thus we must assign a value to the systematic uncertainty to $a^{-1}$ of about 100 MeV which we add in quadrature to a statistical error of about 50 MeV to find

$$a^{-1} \;=\; 1900 \pm 110 \text{ MeV} \quad \text{for} \quad am_q = 0.010. \tag{18}$$

Similarly, we quote

$$a^{-1} \;=\; 1660 \pm 150 \text{ MeV} \quad \text{for} \quad am_q = 0.025. \tag{19}$$

One puzzling feature of the calculation is the very different lattice spacings found here using heavy Wilson quarks and the lattice spacing of the NRQCD group. The numbers are shown in Table XVI, along with all other lattice spacings extracted from this data set of which we are aware: in the table S and W label staggered and Wilson valence quarks, the "force" is from the string tension [22], and the zero quark mass line is from an extrapolation when it is available.

Presently we have calculated only the $^3P_1 - \overline{S}$ mass difference on our quenched configurations (see Table XVII). The lattice spacing given by that measurement is

$$a^{-1} \;=\; 1800 \pm 180 \text{ MeV} \quad \text{for} \quad am_q = \infty. \tag{20}$$

We should point out that for this mass difference, it was necessary to fit the axial-vector propagator to a different range than the pseudoscalar and vector propagators. The latter two take a longer time to "saturate" the signal, and the axial-vector propagator becomes noise dominated after some time. The ranges used are those indicated in Tables XI-XIII.

### C. The Lattice Calculation of $\alpha_s$

The expectation value of the plaquette determines the coupling on the lattice at momentum scale $3.41/a$. However, the fact that we do not know the coefficient in front of the $\mathcal{O}(\alpha_V^3)$ is a source of systematic uncertainty. Thus we invert (9) neglecting higher order terms and estimate our systematic error to be $(1.19 + 0.07 n_f)^2 \alpha_V^3$.

Using the procedure outlined above, we find that, for our dynamical fermion simulations where $\langle \text{Tr } U_{\text{plaq}} \rangle / 3 = 0.5650$ and $0.5644$ for $am_q = 0.010$ and $0.025$ respectively,

$$\alpha_V^{(n_f=2)}(3.41/a) = \begin{cases} 0.179 \pm 0.010 & \text{for } am_q = 0.010 \\ 0.179 \pm 0.010 & \text{for } am_q = 0.025 \end{cases} \quad (21)$$

where the errors quoted are systematic. We do not quote an uncertainty for the trace of the plaquette because it is much smaller than our other errors.

Since we are using the mass splitting in the charmonium sector, three flavors of sea quarks contribute to the theory. There are two methods we know of to convert $\alpha_V$ from $n_f = 2$ to $n_f = 3$. In the first we run $\alpha_V$ down to some low momentum scale ($\sim 500 - 1000$ MeV) with two flavors using the two-loop $\beta$-function,

$$\alpha(q) = \alpha_{\text{I}}(q) + \alpha_{\text{II}}(q) + \mathcal{O}(\alpha^3), \quad (22)$$

where

$$\alpha_{\text{I}}(q) = \frac{\alpha(q_0)}{1 + \alpha(q_0) \beta_0 t}, \quad (23)$$

$$\alpha_{\text{II}}(q) = -[\alpha_{\text{I}}(q)]^2 b_1 \log\left(\frac{\alpha(q_0)}{\alpha_{\text{I}}(q)}\right), \quad (24)$$

with the definitions

$$t = \frac{1}{4\pi} \log\left(\frac{q^2}{q_0^2}\right), \quad \beta_0 = 11 - \frac{2}{3} n_f, \quad \beta_1 = 102 - \frac{38}{3} n_f, \quad \text{and} \quad b_1 = \frac{\beta_1}{4\pi \beta_0}. \quad (25)$$

However, this method is inadequate because we do not know the scale of the strange quark threshold, and because our perturbative expression (22) fails as $\alpha_V$ becomes large.

The second method is to use our quenched data to give us the strong coupling for $n_f = 0$. From $\langle \text{Tr } U_{\text{plaq}} \rangle / 3 = 0.59367$ we find that $\alpha_V^{(n_f=0)}(3.41/a) = 0.152(4)$. Since the inverse coupling is nearly linear in $n_f$, we can extrapolate to $n_f = 3$. However, since the lattice spacings are different for the dynamical fermion simulations and the quenched simulation, we must run one of the couplings to the scale at which the other is known using the two-loop $\beta$-function (22).

We need only one value for $\alpha_V^{(n_f=2)}$, so for the time being we shall use the lattice spacing from the $am_q = 0.010$ staggered fermion data. Thus we have the strong coupling measured with either zero or two flavors of sea quarks:

$$\alpha_V^{(n_f=0)}(6.14 \text{ GeV}) = 0.152 \pm 0.005 \pm 0.005$$
$$\alpha_V^{(n_f=2)}(6.48 \text{ GeV}) = 0.179 \pm 0.004 \pm 0.010, \tag{26}$$

where the first errors quoted are propagated from the uncertainty in the lattice spacing, and the second errors are our estimation of the systematic uncertainties resulting from the perturbative calculations. We choose to run the 0 flavor coupling to 6.48 GeV, giving

$$\alpha_V(6.48 \text{ GeV}) = \begin{cases} 0.152 \pm 0.006 & \text{for } n_f = 0 \\ 0.179 \pm 0.011 & \text{for } n_f = 2 \end{cases} \tag{27}$$

Here the statistical and systematic uncertainties have been combined in quadrature. If we had chosen to perform the extrapolation to three flavors at 6.14 GeV instead, there would be no change in our determination of $\alpha_V$ at any scale.

We may linearly extrapolate either $1/\alpha$ or $\alpha$ from zero to three flavors. (In first order $1/\alpha$ depends linearly on $n_f$.) Extrapolating in $1/\alpha$ gives $\alpha_V^{(n_f=3)}(6.48 \text{ GeV}) = 0.198 \pm 0.020$. However, linear extrapolation in $\alpha$ has smaller errors,

$$\alpha_V^{(n_f=3)}(6.48 \text{ GeV}) = 0.194 \pm 0.017, \tag{28}$$

and we will use it below.

### D. Lattice Prescription to Minimal Subtraction Schemes

In order to compare our result with experimental determinations of the strong coupling constant, we convert from our lattice definition of $\alpha_V$ (28) to the coupling defined in the modified minimal subtraction scheme, $\alpha_{\overline{MS}}$. The two couplings are related perturbatively by

$$\alpha_{\overline{MS}}(q) = \alpha_V(qe^{5/6})\left(1 + \frac{2\alpha_V}{\pi} + \mathcal{O}(\alpha_V^2)\right). \tag{29}$$

Inserting (28) into the above expression gives

$$\alpha_{\overline{MS}}^{(n_f=3)}(2.82 \text{ GeV}) = 0.218 \pm 0.021 \pm 0.007, \tag{30}$$

where the first error is propagated from the error in $\alpha_V$, and the second is a systematic error of $\alpha_V^3$ assumed from the perturbative expression (29).

Once we have converted to the $\overline{MS}$ subtraction scheme, we can include the 3-loop term in the $\beta$-function (22), (see [23]):

$$\alpha_{\text{III}}(q) = [\alpha_{\text{I}}(q)]^3 \left\{ b_1^2 \log\left(\frac{\alpha(q_0)}{\alpha_{\text{I}}(q)}\right)\left[\log\left(\frac{\alpha(q_0)}{\alpha_{\text{I}}(q)}\right) - 1\right] - (b_1^2 - b_2)\left(1 - \frac{\alpha(q_0)}{\alpha_{\text{I}}(q)}\right)\right\}, \tag{31}$$

where

$$\beta_2 = \frac{1}{2}\left(2857 - \frac{5033}{9}n_f + \frac{325}{27}n_f^2\right), \text{ and } b_2 = \frac{\beta_2}{(4\pi)^2\beta_0}. \tag{32}$$

Thus our error from running the coupling with this $\beta$-function should be of order $\alpha^4$, which is always smaller than the present errors.

Rodrigo and Santamaria [23] show that the heavy quark thresholds in $\overline{MS}$ prescription are at the mass of the quark. When we cross the threshold, we increment $n_f$ by one. Starting with

$$\alpha_{\overline{MS}}^{(n_f=3)}(2.82 \text{ GeV}) = 0.218 \pm 0.022, \tag{33}$$

we run down to the charm quark mass with three flavors, and change to four. Then we can run to the bottom quark mass and change from four flavors to five. Finally, we can run the coupling up to the $Z$ mass with five flavors. The results are as follows:

$$\alpha_{\overline{MS}}^{(n_f=3,4)}(q = 1.5 \text{ GeV} \simeq m_c) = 0.287(38) \tag{34}$$

$$\alpha_{\overline{MS}}^{(n_f=4,5)}(q = 4.5 \text{ GeV} \simeq m_b) = 0.195(17) \tag{35}$$

$$\alpha_{\overline{MS}}^{(n_f=5)}(q = 91.2 \text{ GeV} = M_Z) = 0.111(6). \tag{36}$$

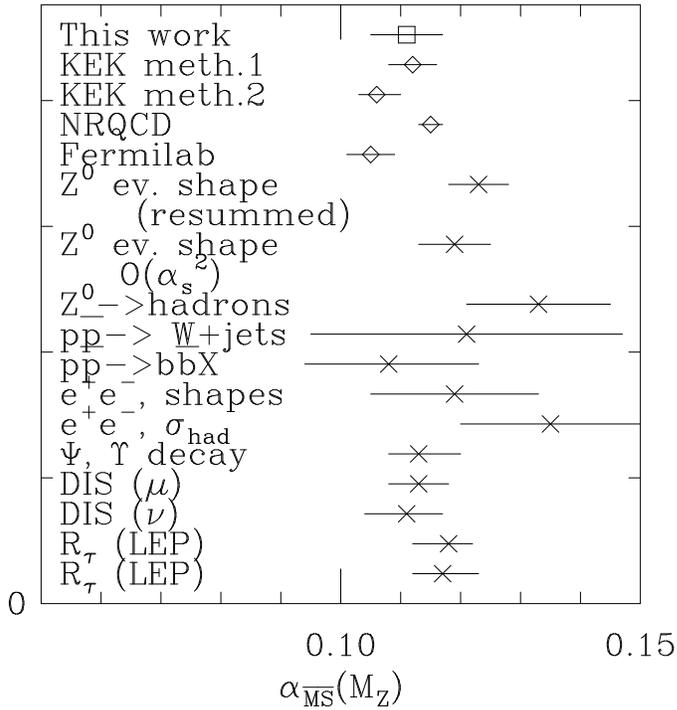

FIG. 10. Recent measurements of the strong coupling constant at the Z-mass. The square marks our result. Diamonds represent other lattice calculations, and crosses are experimental results.

We now compare our number to other lattice determinations of the strong coupling constant and to experiment. It has become conventional to make these comparisons at the Z pole. Fig. 10 shows a compilation of recent lattice and experimental results, from Ref. [24].

It is interesting that, although the lattice spacings using Wilson quarks are significantly lower than those found on the same configurations using non-relativistic quarks, the determinations of the strong coupling are in good agreement [4].

The world average [25] for the strong coupling at the $Z$ boson mass is $\alpha_{\overline{MS}}(M_Z) = 0.117(5)$. Our number (36) is in agreement with this average.

Running to the $Z$ mass somewhat artificially compresses all the uncertainties of a low-$Q^2$ calculation (the $g = 0$ fixed point is at infinite $Q$), and it might be more revealing to compare the lattice prediction to an experimental measurement at a low energy scale. The CLEO collaboration [26] gives the strong coupling at the $\tau$ lepton mass as $\alpha_{\overline{MS}}(M_\tau) = 0.309(24)$. By running the $\overline{MS}$ coupling with $n_f = 3$ down to this scale we find a lattice prediction of

$$\alpha_{\overline{MS}}^{(n_f=3)}(M_\tau = 1.777 \text{ GeV}) = 0.265(32), \quad (37)$$

which is in loose agreement with CLEO.

Following the same procedure using the data with sea quarks of mass $am_q = 0.025$ rather than 0.010 gives the following numbers:

$$\alpha_{\overline{MS}}^{(n_f=3)}(M_\tau) = 0.245(29) \quad (38)$$

$$\alpha_{\overline{MS}}^{(n_f=5)}(M_Z) = 0.107(5). \quad (39)$$

Since the data seems to be less noisy for the $am_q = 0.010$ data than the $am_q = 0.025$ data, we will use the former in quoting our final results, (36) and (37).

## V. CHARM SPECTROSCOPY

Although there is already evidence that Wilson quarks do not calculate the hyperfine structure correctly and it is known that heavy Wilson fermions have an anomalously small lattice artifact magnetic moment [27], we attempt to calculate the masses of the $L = 0$ and $L = 1$ $D$-meson states.

In Section IV A we estimated the hopping parameter of the charm quark, $\kappa_{\text{charm}}$, in order to calculate the S-P mass splitting of charmonium. We could only guess roughly at $\kappa_{\text{charm}}$ since we did not know the lattice spacing. However, the S-P splitting is very weakly dependent on the quark mass, so we could determine it without knowing $\kappa_{\text{charm}}$. On the other hand, while the mass splitting is independent of quark mass, the masses of the charm states are not. Thus having found the lattice spacing, we can find $\kappa_{\text{charm}}$ accurately and compute the mass spectrum of the charmed meson system.

As we mentioned in Sec. IV A we must shift the lattice meson mass using Eqn. (13). In what follows we perform a jackknife analysis, dividing our 100 configurations into 10 sets of 90 sequential lattices for the dynamical fermion simulations and dividing our 79 quenched lattices into 7 sets of 68. We find the best fits to the vector meson at our three lowest values of the hopping parameter and correct the heavy quark mass using Eqns. (13)-(15). Then we interpolate linearly between these lattice masses to find the hopping parameter for which $aM_\psi = a(3.1 \text{ GeV})$. The error in the lattice spacing gives us a range for $\kappa_{\text{charm}}$.

For the $am_q = 0.010$ dynamical fermion data set, we find $\kappa_{\text{charm}} = 0.134 \pm 0.002$, while for $am_q = 0.025$ we find $\kappa_{\text{charm}} = 0.129 \pm 0.003$, and for the quenched data set $\kappa_{\text{charm}} = 0.128 \pm 0.003$.

We can compute the masses of the $D$ mesons by looking at our heavy-light meson states, extrapolating the light quark mass to zero, i.e. $\kappa_{\text{light}} \to \kappa_{\text{crit}}$, and interpolating the heavy quark mass to the charm mass, $\kappa_{\text{heavy}} \to \kappa_{\text{charm}}$. For the $n_f = 2$ data we divide our six flavors of valence quarks into heavy and light as follows: $\kappa_{\text{heavy}} \in \{\kappa_1, \kappa_2, \kappa_3\} = \{0.1320, 0.1410, 0.1525\}$ and $\kappa_{\text{light}} \in \{\kappa_4, \kappa_5, \kappa_6\} = \{0.1565, 0.1585, 0.1600\}$. The quenched data set has two heavy kappas and three light ones: $\kappa_{\text{heavy}} \in \{\kappa_0, \kappa_1\} = \{0.1300, 0.1450\}$ and $\kappa_{\text{light}} \in \{\kappa_2, \kappa_3, \kappa_4\} = \{0.1520, 0.1540, 0.1550\}$. Using the jackknife procedure mentioned above, for each $\kappa_{\text{heavy}}$ separately, we use the best fit ranges to the three mesons with different light quarks and extrapolate to $\kappa_{\text{light}} = \kappa_{\text{crit}}$. Having done this for the values of $\kappa_{\text{heavy}}$ we can interpolate to $\kappa_{\text{heavy}} = \kappa_{\text{charm}}$.

We must shift the meson mass as with the $J/\psi$, but since the $D$ mesons have only one heavy quark we replace Eqn. (13) by

$$M_D a = \mu - m_1 a + m_2 a, \qquad (40)$$

with the identities (14) and (15). The results are summarized in Tables XVIII, XIX, and XX. Our results are consistently lower than those from experiment [21,28]. Using the cubic group representations we found complete degeneracy among the P-wave states, in disagreement with experiment.

We also calculate the mass differences

$$M(D^*) - M(D^0), \qquad (41)$$

and

$$M(D_1) - \frac{1}{4}\big[M(D^0) + 3M(D^*)\big], \qquad (42)$$

for the $am_q = 0.010$ dynamical fermion and the quenched data sets. The other dynamical fermion data set yielded consistent results, but with much larger errors. See Tables XXI and XXII for results.

## VI. CONCLUSIONS

This simulation had a number of inadequacies. First, our P-wave data are much noisier than S-wave spectroscopy. Lattice simulations with light fermions need better interpolating fields for P-wave states. It is much easier to explore large classes of trial wave functions and to find optimal ones when one has one or more heavy or static quarks (cf. Refs. [29] and [30]), and so heavy Wilson simulations are presently just not competitive with simulations with one static or nonrelativistic quark from the point of view of statistical uncertainties. Second, heavy Wilson fermions exhibit known lattice artefacts which should be corrected in future simulations through the use of improved actions. NRQCD remains the method of choice for heavy quark systems. However, the large scaling violations between relativistic fermion simulations and NRQCD need to be understood. Nevertheless, it is possible to make a determination of the strong coupling constant with an uncertainty comparable with other recent measurements, and which does not have the associated uncertainty induced by the absence of sea quarks.

The final result for the strong coupling constant from this work is $\alpha_{\overline{MS}}^{(n_f=5)}(M_Z) = 0.111(6)$, or $\alpha_{\overline{MS}}^{(n_f=3)}(M_\tau) = 0.265(32)$.

## ACKNOWLEDGMENTS

This work was supported by the U.S. Department of Energy under grants DE-FG05-85ER250000, DE-FG05-92ER40742, DE-FG02–92ER–40672, DE-FG03–90ER–40546, DE-FG02–91ER–40661, and National Science Foundation Grants No. NSF–PHY91–01853 . Simulations were carried out at the Supercomputer Computations Research Institute at Florida State University, at the San Diego Supercomputer Center, and at Indiana University.

We would like to thank our colleagues on the MILC collaboration for their help. Also, TD would like to extend thanks to the participants at a workshop in Santa Fe, N. M., especially A. El Khadra and J. Shigemitsu, for helpful discussions. So would SC and UMH to John Sloan.

# TABLES

TABLE I. Best range fits to $M(^3P_1)$ for $am_q = 0.010$ using the axial vector current.

| $i$ | $j$ | $(\kappa_i + \kappa_j)/2$ | $aM(^3P_1)$ | error | $\chi^2/N_{DF}$ | conf. lvl. | $t_i$ | $t_f$ |
|---|---|---|---|---|---|---|---|---|
| 1 | 1 | 0.1320 | 1.724 | 0.008 | 3.60/4 | 0.462 | 5 | 10 |
| 2 | 1 | 0.1365 | 1.562 | 0.008 | 3.38/4 | 0.497 | 5 | 10 |
| 2 | 2 | 0.1410 | 1.398 | 0.007 | 3.41/4 | 0.491 | 5 | 10 |
| 3 | 1 | 0.1422 | 1.360 | 0.008 | 4.22/4 | 0.377 | 5 | 10 |
| 3 | 2 | 0.1467 | 1.191 | 0.008 | 4.72/4 | 0.317 | 5 | 10 |
| 3 | 3 | 0.1525 | 0.976 | 0.010 | 5.12/4 | 0.275 | 5 | 10 |
| 4 | 1 | 0.1442 | 1.292 | 0.009 | 3.66/4 | 0.454 | 5 | 10 |
| 4 | 2 | 0.1487 | 1.122 | 0.009 | 4.75/4 | 0.314 | 5 | 10 |
| 4 | 3 | 0.1545 | 0.900 | 0.010 | 5.14/4 | 0.274 | 5 | 10 |
| 4 | 4 | 0.1565 | 0.820 | 0.012 | 5.32/4 | 0.256 | 5 | 10 |
| 5 | 1 | 0.1452 | 1.259 | 0.010 | 1.97/4 | 0.742 | 5 | 10 |
| 5 | 2 | 0.1497 | 1.086 | 0.010 | 3.87/4 | 0.424 | 5 | 10 |
| 5 | 3 | 0.1555 | 0.861 | 0.011 | 4.89/4 | 0.299 | 5 | 10 |
| 5 | 4 | 0.1575 | 0.778 | 0.013 | 4.19/4 | 0.381 | 5 | 10 |
| 5 | 5 | 0.1585 | 0.736 | 0.015 | 2.50/4 | 0.644 | 5 | 10 |
| 6 | 1 | 0.1460 | 1.238 | 0.014 | 2.52/4 | 0.642 | 5 | 10 |
| 6 | 2 | 0.1505 | 1.060 | 0.013 | 5.23/4 | 0.264 | 5 | 10 |
| 6 | 3 | 0.1562 | 0.834 | 0.014 | 5.07/4 | 0.280 | 5 | 10 |
| 6 | 4 | 0.1583 | 0.753 | 0.016 | 2.57/4 | 0.632 | 5 | 10 |
| 6 | 5 | 0.1593 | 0.749 | 0.010 | 5.70/6 | 0.458 | 3 | 10 |
| 6 | 6 | 0.1600 | 0.708 | 0.013 | 2.52/6 | 0.866 | 3 | 10 |

TABLE II. Best range fits to $M(^3P_2)$ for $am_q = 0.010$ using the cubic group state $E_a$.

| $i$ | $j$ | $(\kappa_i + \kappa_j)/2$ | $aM(^3P_2)$ | error | $\chi^2/N_{DF}$ | conf. lvl. | $t_i$ | $t_f$ |
|---|---|---|---|---|---|---|---|---|
| 1 | 1 | 0.1320 | 1.706 | 0.013 | 1.14/4 | 0.888 | 5 | 10 |
| 2 | 1 | 0.1365 | 1.537 | 0.015 | 0.25/4 | 0.993 | 5 | 10 |
| 2 | 2 | 0.1410 | 1.380 | 0.015 | 0.61/4 | 0.962 | 5 | 10 |
| 3 | 1 | 0.1422 | 1.348 | 0.019 | 0.66/4 | 0.956 | 5 | 10 |
| 3 | 2 | 0.1467 | 1.189 | 0.020 | 0.45/4 | 0.979 | 5 | 10 |
| 3 | 3 | 0.1525 | 0.988 | 0.025 | 0.14/4 | 0.998 | 5 | 10 |
| 4 | 1 | 0.1442 | 1.295 | 0.023 | 3.63/4 | 0.458 | 5 | 10 |
| 4 | 2 | 0.1487 | 1.141 | 0.024 | 3.39/4 | 0.495 | 5 | 10 |
| 4 | 3 | 0.1545 | 0.956 | 0.031 | 2.25/4 | 0.691 | 5 | 10 |
| 4 | 4 | 0.1565 | 0.897 | 0.038 | 1.08/4 | 0.898 | 5 | 10 |
| 5 | 1 | 0.1452 | 1.267 | 0.029 | 6.60/4 | 0.159 | 5 | 10 |
| 5 | 2 | 0.1497 | 1.124 | 0.030 | 7.30/4 | 0.121 | 5 | 10 |
| 5 | 3 | 0.1555 | 0.966 | 0.037 | 6.90/4 | 0.142 | 5 | 10 |
| 5 | 4 | 0.1575 | 0.975 | 0.038 | 7.32/5 | 0.198 | 4 | 10 |
| 5 | 5 | 0.1585 | 0.871 | 0.055 | 4.27/4 | 0.371 | 5 | 10 |
| 6 | 1 | 0.1460 | 1.224 | 0.038 | 6.99/4 | 0.136 | 5 | 10 |
| 6 | 2 | 0.1505 | 1.093 | 0.040 | 9.69/4 | 0.046 | 5 | 10 |
| 6 | 3 | 0.1562 | 1.031 | 0.043 | 14.88/5 | 0.011 | 4 | 10 |
| 6 | 4 | 0.1583 | 1.008 | 0.050 | 14.20/5 | 0.014 | 4 | 10 |
| 6 | 5 | 0.1593 | 1.003 | 0.062 | 15.39/5 | 0.009 | 4 | 10 |
| 6 | 6 | 0.1600 | 0.973 | 0.082 | 15.61/5 | 0.008 | 4 | 10 |

TABLE III. Best range fits to $M(^3P_1)$ for $am_q = 0.010$ using the cubic group state $T_{1a}$.

| $i$ | $j$ | $(\kappa_i + \kappa_j)/2$ | $aM(^3P_1)$ | error | $\chi^2/N_{DF}$ | conf. lvl. | $t_i$ | $t_f$ |
|---|---|---|---|---|---|---|---|---|
| 1 | 1 | 0.1320 | 1.709 | 0.012 | 0.99/4 | 0.912 | 5 | 10 |
| 2 | 1 | 0.1365 | 1.538 | 0.014 | 0.56/4 | 0.967 | 5 | 10 |
| 2 | 2 | 0.1410 | 1.384 | 0.014 | 0.61/4 | 0.962 | 5 | 10 |
| 3 | 1 | 0.1422 | 1.336 | 0.017 | 1.86/4 | 0.762 | 5 | 10 |
| 3 | 2 | 0.1467 | 1.178 | 0.016 | 1.46/4 | 0.833 | 5 | 10 |
| 3 | 3 | 0.1525 | 0.980 | 0.018 | 0.66/4 | 0.956 | 5 | 10 |
| 4 | 1 | 0.1442 | 1.297 | 0.027 | 0.87/3 | 0.832 | 6 | 10 |
| 4 | 2 | 0.1487 | 1.111 | 0.019 | 3.86/4 | 0.426 | 5 | 10 |
| 4 | 3 | 0.1545 | 0.911 | 0.021 | 3.79/4 | 0.435 | 5 | 10 |
| 4 | 4 | 0.1565 | 0.850 | 0.025 | 4.91/4 | 0.296 | 5 | 10 |
| 5 | 1 | 0.1452 | 1.279 | 0.036 | 1.03/3 | 0.795 | 6 | 10 |
| 5 | 2 | 0.1497 | 1.119 | 0.035 | 1.71/3 | 0.635 | 6 | 10 |
| 5 | 3 | 0.1555 | 0.878 | 0.026 | 6.35/4 | 0.174 | 5 | 10 |
| 5 | 4 | 0.1575 | 0.812 | 0.032 | 8.28/4 | 0.082 | 5 | 10 |
| 5 | 5 | 0.1585 | 0.781 | 0.040 | 9.77/4 | 0.044 | 5 | 10 |
| 6 | 1 | 0.1460 | 1.281 | 0.057 | 1.21/3 | 0.751 | 6 | 10 |
| 6 | 2 | 0.1505 | 1.052 | 0.036 | 5.41/4 | 0.248 | 5 | 10 |
| 6 | 3 | 0.1562 | 0.846 | 0.038 | 7.67/4 | 0.104 | 5 | 10 |
| 6 | 4 | 0.1583 | 0.769 | 0.044 | 9.27/4 | 0.055 | 5 | 10 |
| 6 | 5 | 0.1593 | 0.712 | 0.054 | 9.02/4 | 0.061 | 5 | 10 |
| 6 | 6 | 0.1600 | 0.659 | 0.077 | 5.90/4 | 0.207 | 5 | 10 |

TABLE IV. Best range fits to $M(^3P_0)$ for $am_q = 0.010$ using the cubic group state $A_1$.

| $i$ | $j$ | $(\kappa_i + \kappa_j)/2$ | $aM(^3P_0)$ | error | $\chi^2/N_{DF}$ | conf. lvl. | $t_i$ | $t_f$ |
|---|---|---|---|---|---|---|---|---|
| 1 | 1 | 0.1320 | 1.712 | 0.013 | 2.31/4 | 0.679 | 5 | 10 |
| 2 | 1 | 0.1365 | 1.541 | 0.014 | 1.21/4 | 0.876 | 5 | 10 |
| 2 | 2 | 0.1410 | 1.387 | 0.014 | 1.54/4 | 0.820 | 5 | 10 |
| 3 | 1 | 0.1422 | 1.335 | 0.017 | 1.65/4 | 0.800 | 5 | 10 |
| 3 | 2 | 0.1467 | 1.175 | 0.017 | 1.52/4 | 0.824 | 5 | 10 |
| 3 | 3 | 0.1525 | 0.973 | 0.020 | 1.19/4 | 0.880 | 5 | 10 |
| 4 | 1 | 0.1442 | 1.267 | 0.021 | 2.62/4 | 0.624 | 5 | 10 |
| 4 | 2 | 0.1487 | 1.104 | 0.020 | 2.21/4 | 0.696 | 5 | 10 |
| 4 | 3 | 0.1545 | 0.897 | 0.023 | 1.36/4 | 0.852 | 5 | 10 |
| 4 | 4 | 0.1565 | 0.834 | 0.027 | 1.11/4 | 0.892 | 5 | 10 |
| 5 | 1 | 0.1452 | 1.237 | 0.025 | 3.16/4 | 0.532 | 5 | 10 |
| 5 | 2 | 0.1497 | 1.072 | 0.023 | 2.69/4 | 0.611 | 5 | 10 |
| 5 | 3 | 0.1555 | 0.859 | 0.026 | 1.79/4 | 0.774 | 5 | 10 |
| 5 | 4 | 0.1575 | 0.793 | 0.032 | 1.63/4 | 0.803 | 5 | 10 |
| 5 | 5 | 0.1585 | 0.876 | 0.087 | 0.07/2 | 0.967 | 7 | 10 |
| 6 | 1 | 0.1460 | 1.208 | 0.035 | 2.39/4 | 0.665 | 5 | 10 |
| 6 | 2 | 0.1505 | 1.044 | 0.032 | 3.10/4 | 0.541 | 5 | 10 |
| 6 | 3 | 0.1562 | 0.844 | 0.048 | 4.76/3 | 0.191 | 6 | 10 |
| 6 | 4 | 0.1583 | 0.836 | 0.038 | 10.82/5 | 0.055 | 4 | 10 |
| 6 | 5 | 0.1593 | 0.728 | 0.063 | 2.50/3 | 0.475 | 6 | 10 |
| 6 | 6 | 0.1600 | 0.826 | 0.062 | 4.45/4 | 0.349 | 5 | 10 |

TABLE V. Best range fits to $M(^1P_1)$ for $am_q = 0.010$ using the cubic group state $^1P_1$.

| $i$ | $j$ | $(\kappa_i + \kappa_j)/2$ | $aM(^1P_1)$ | error | $\chi^2/N_{DF}$ | conf. lvl. | $t_i$ | $t_f$ |
|---|---|---|---|---|---|---|---|---|
| 1 | 1 | 0.1320 | 1.713 | 0.012 | 1.81/4 | 0.770 | 5 | 10 |
| 2 | 1 | 0.1365 | 1.543 | 0.013 | 0.77/4 | 0.942 | 5 | 10 |
| 2 | 2 | 0.1410 | 1.389 | 0.013 | 1.20/4 | 0.878 | 5 | 10 |
| 3 | 1 | 0.1422 | 1.346 | 0.015 | 0.59/4 | 0.964 | 5 | 10 |
| 3 | 2 | 0.1467 | 1.189 | 0.014 | 0.41/4 | 0.982 | 5 | 10 |
| 3 | 3 | 0.1525 | 0.993 | 0.016 | 0.87/4 | 0.929 | 5 | 10 |
| 4 | 1 | 0.1442 | 1.282 | 0.017 | 1.41/4 | 0.843 | 5 | 10 |
| 4 | 2 | 0.1487 | 1.124 | 0.016 | 0.99/4 | 0.911 | 5 | 10 |
| 4 | 3 | 0.1545 | 0.926 | 0.017 | 0.47/4 | 0.976 | 5 | 10 |
| 4 | 4 | 0.1565 | 0.865 | 0.020 | 0.68/4 | 0.953 | 5 | 10 |
| 5 | 1 | 0.1452 | 1.251 | 0.020 | 2.52/4 | 0.641 | 5 | 10 |
| 5 | 2 | 0.1497 | 1.095 | 0.019 | 2.40/4 | 0.662 | 5 | 10 |
| 5 | 3 | 0.1555 | 0.894 | 0.019 | 1.77/4 | 0.778 | 5 | 10 |
| 5 | 4 | 0.1575 | 0.831 | 0.022 | 1.90/4 | 0.753 | 5 | 10 |
| 5 | 5 | 0.1585 | 0.806 | 0.027 | 2.65/4 | 0.618 | 5 | 10 |
| 6 | 1 | 0.1460 | 1.233 | 0.026 | 3.83/4 | 0.430 | 5 | 10 |
| 6 | 2 | 0.1505 | 1.077 | 0.024 | 4.53/4 | 0.339 | 5 | 10 |
| 6 | 3 | 0.1562 | 0.872 | 0.025 | 4.91/4 | 0.297 | 5 | 10 |
| 6 | 4 | 0.1583 | 0.804 | 0.031 | 4.06/4 | 0.398 | 5 | 10 |
| 6 | 5 | 0.1593 | 0.777 | 0.039 | 3.20/4 | 0.525 | 5 | 10 |
| 6 | 6 | 0.1600 | 0.778 | 0.059 | 2.35/4 | 0.672 | 5 | 10 |

TABLE VI. Best range fits to $M(^3P_1)$ for $am_q = 0.025$ using an axial vector current.

| $i$ | $j$ | $(\kappa_i + \kappa_j)/2$ | $aM(^3P_1)$ | error | $\chi^2/N_{DF}$ | conf. lvl. | $t_i$ | $t_f$ |
|---|---|---|---|---|---|---|---|---|
| 1 | 1 | 0.1320 | 1.695 | 0.016 | 15.91/2 | 0.000 | 9 | 12 |
| 2 | 1 | 0.1365 | 1.543 | 0.016 | 13.63/2 | 0.001 | 9 | 12 |
| 2 | 2 | 0.1410 | 1.430 | 0.012 | 1.47/4 | 0.832 | 7 | 12 |
| 3 | 1 | 0.1422 | 1.337 | 0.021 | 21.97/2 | 0.000 | 9 | 12 |
| 3 | 2 | 0.1467 | 1.223 | 0.014 | 1.38/4 | 0.847 | 7 | 12 |
| 3 | 3 | 0.1525 | 1.000 | 0.018 | 1.82/4 | 0.768 | 7 | 12 |
| 4 | 1 | 0.1442 | 1.317 | 0.017 | 20.52/4 | 0.000 | 7 | 12 |
| 4 | 2 | 0.1487 | 1.145 | 0.015 | 2.06/4 | 0.725 | 7 | 12 |
| 4 | 3 | 0.1545 | 0.911 | 0.021 | 2.13/4 | 0.712 | 7 | 12 |
| 4 | 4 | 0.1565 | 0.816 | 0.026 | 2.01/4 | 0.734 | 7 | 12 |
| 5 | 1 | 0.1452 | 1.275 | 0.020 | 12.19/4 | 0.016 | 7 | 12 |
| 5 | 2 | 0.1497 | 1.099 | 0.018 | 2.59/4 | 0.629 | 7 | 12 |
| 5 | 3 | 0.1555 | 0.861 | 0.024 | 2.50/4 | 0.645 | 7 | 12 |
| 5 | 4 | 0.1575 | 0.762 | 0.029 | 2.39/4 | 0.665 | 7 | 12 |
| 5 | 5 | 0.1585 | 0.706 | 0.036 | 3.04/4 | 0.551 | 7 | 12 |
| 6 | 1 | 0.1460 | 1.236 | 0.025 | 5.82/4 | 0.213 | 7 | 12 |
| 6 | 2 | 0.1505 | 1.061 | 0.023 | 4.26/4 | 0.372 | 7 | 12 |
| 6 | 3 | 0.1562 | 0.822 | 0.029 | 3.21/4 | 0.524 | 7 | 12 |
| 6 | 4 | 0.1583 | 0.720 | 0.036 | 3.14/4 | 0.534 | 7 | 12 |
| 6 | 5 | 0.1593 | 0.772 | 0.081 | 0.54/3 | 0.910 | 8 | 12 |
| 6 | 6 | 0.1600 | 0.797 | 0.148 | 0.13/3 | 0.988 | 8 | 12 |

TABLE VII. Best range fits to $M(^3P_2)$ for $am_q = 0.025$ using the cubic group $E_a$ state.

| $i$ | $j$ | $(\kappa_i + \kappa_j)/2$ | $aM(^3P_2)$ | error | $\chi^2/N_{DF}$ | conf. lvl. | $t_i$ | $t_f$ |
|---|---|---|---|---|---|---|---|---|
| 1 | 1 | 0.1320 | 1.710 | 0.026 | 1.97/3 | 0.579 | 8 | 12 |
| 2 | 1 | 0.1365 | 1.557 | 0.029 | 3.68/3 | 0.298 | 8 | 12 |
| 2 | 2 | 0.1410 | 1.392 | 0.029 | 4.09/3 | 0.252 | 8 | 12 |
| 3 | 1 | 0.1422 | 1.388 | 0.033 | 8.51/4 | 0.075 | 7 | 12 |
| 3 | 2 | 0.1467 | 1.219 | 0.034 | 7.83/4 | 0.098 | 7 | 12 |
| 3 | 3 | 0.1525 | 1.018 | 0.045 | 7.31/4 | 0.120 | 7 | 12 |
| 4 | 1 | 0.1442 | 1.363 | 0.063 | 9.44/3 | 0.024 | 8 | 12 |
| 4 | 2 | 0.1487 | 1.124 | 0.084 | 2.79/2 | 0.248 | 9 | 12 |
| 4 | 3 | 0.1545 | 1.035 | 0.070 | 8.04/4 | 0.090 | 7 | 12 |
| 4 | 4 | 0.1565 | 7.799 | 0.000 | 22.25/2 | 0.000 | 9 | 12 |
| 5 | 1 | 0.1452 | 1.344 | 0.092 | 2.64/2 | 0.267 | 9 | 12 |
| 5 | 2 | 0.1497 | 1.242 | 0.033 | 3.89/2 | 0.143 | 9 | 12 |
| 5 | 3 | 0.1555 | 0.938 | 0.027 | 8.95/5 | 0.111 | 4 | 10 |
| 5 | 4 | 0.1575 | 0.879 | 0.034 | 11.12/5 | 0.049 | 4 | 10 |
| 5 | 5 | 0.1585 | 0.866 | 0.042 | 11.74/5 | 0.038 | 4 | 10 |
| 6 | 1 | 0.1460 | 1.085 | 0.114 | 0.53/2 | 0.768 | 9 | 12 |
| 6 | 2 | 0.1505 | 0.888 | 0.077 | 0.81/3 | 0.847 | 8 | 12 |
| 6 | 3 | 0.1562 | 0.908 | 0.035 | 4.35/5 | 0.500 | 4 | 10 |
| 6 | 4 | 0.1583 | 0.849 | 0.042 | 5.53/5 | 0.354 | 4 | 10 |
| 6 | 5 | 0.1593 | 0.837 | 0.050 | 6.20/5 | 0.287 | 4 | 10 |
| 6 | 6 | 0.1600 | 0.863 | 0.063 | 5.80/5 | 0.327 | 4 | 10 |

TABLE VIII. Best range fits to $M(^3P_1)$ for $am_q = 0.025$ using the cubic group $T_{1a}$ state.

| $i$ | $j$ | $(\kappa_i + \kappa_j)/2$ | $aM(^3P_1)$ | error | $\chi^2/N_{DF}$ | conf. lvl. | $t_i$ | $t_f$ |
|---|---|---|---|---|---|---|---|---|
| 1 | 1 | 0.1320 | 1.714 | 0.028 | 0.82/3 | 0.846 | 8 | 12 |
| 2 | 1 | 0.1365 | 1.563 | 0.031 | 1.60/3 | 0.660 | 8 | 12 |
| 2 | 2 | 0.1410 | 1.398 | 0.031 | 1.69/3 | 0.639 | 8 | 12 |
| 3 | 1 | 0.1422 | 1.370 | 0.038 | 3.28/3 | 0.351 | 8 | 12 |
| 3 | 2 | 0.1467 | 1.208 | 0.039 | 3.36/3 | 0.339 | 8 | 12 |
| 3 | 3 | 0.1525 | 0.994 | 0.050 | 2.51/3 | 0.473 | 8 | 12 |
| 4 | 1 | 0.1442 | 1.303 | 0.046 | 4.18/3 | 0.242 | 8 | 12 |
| 4 | 2 | 0.1487 | 1.152 | 0.054 | 3.99/3 | 0.263 | 8 | 12 |
| 4 | 3 | 0.1545 | 1.077 | 0.073 | 4.36/4 | 0.360 | 7 | 12 |
| 4 | 4 | 0.1565 | 0.956 | 0.041 | 2.41/2 | 0.300 | 9 | 12 |
| 5 | 1 | 0.1452 | 1.284 | 0.067 | 4.29/3 | 0.232 | 8 | 12 |
| 5 | 2 | 0.1497 | 1.260 | 0.063 | 6.21/4 | 0.184 | 7 | 12 |
| 5 | 3 | 0.1555 | 0.985 | 0.020 | 7.10/5 | 0.213 | 4 | 10 |
| 5 | 4 | 0.1575 | 0.921 | 0.023 | 8.47/5 | 0.132 | 4 | 10 |
| 5 | 5 | 0.1585 | 0.894 | 0.027 | 9.67/5 | 0.085 | 4 | 10 |
| 6 | 1 | 0.1460 | 1.433 | 0.112 | 5.90/4 | 0.207 | 7 | 12 |
| 6 | 2 | 0.1505 | 1.331 | 0.113 | 5.95/4 | 0.203 | 7 | 12 |
| 6 | 3 | 0.1562 | 0.951 | 0.025 | 10.12/5 | 0.072 | 4 | 10 |
| 6 | 4 | 0.1583 | 0.889 | 0.029 | 12.36/5 | 0.030 | 4 | 10 |
| 6 | 5 | 0.1593 | 0.861 | 0.035 | 14.33/5 | 0.014 | 4 | 10 |
| 6 | 6 | 0.1600 | 0.837 | 0.044 | 14.27/5 | 0.014 | 4 | 10 |

TABLE IX. Best range fits to $M(^3P_0)$ for $am_q = 0.025$ using the cubic group $A_1$ state.

| $i$ | $j$ | $(\kappa_i + \kappa_j)/2$ | $aM(^3P_0)$ | error | $\chi^2/N_{DF}$ | conf. lvl. | $t_i$ | $t_f$ |
|---|---|---|---|---|---|---|---|---|
| 1 | 1 | 0.1320 | 1.708 | 0.030 | 0.28/3 | 0.964 | 8 | 12 |
| 2 | 1 | 0.1365 | 1.550 | 0.033 | 0.19/3 | 0.979 | 8 | 12 |
| 2 | 2 | 0.1410 | 1.375 | 0.034 | 0.07/3 | 0.996 | 8 | 12 |
| 3 | 1 | 0.1422 | 1.338 | 0.041 | 1.65/3 | 0.649 | 8 | 12 |
| 3 | 2 | 0.1467 | 1.154 | 0.039 | 1.63/3 | 0.654 | 8 | 12 |
| 3 | 3 | 0.1525 | 0.876 | 0.040 | 1.03/3 | 0.794 | 8 | 12 |
| 4 | 1 | 0.1442 | 1.242 | 0.049 | 4.98/3 | 0.173 | 8 | 12 |
| 4 | 2 | 0.1487 | 1.057 | 0.046 | 5.35/3 | 0.148 | 8 | 12 |
| 4 | 3 | 0.1545 | 0.776 | 0.047 | 3.43/3 | 0.330 | 8 | 12 |
| 4 | 4 | 0.1565 | 0.715 | 0.077 | 1.83/2 | 0.401 | 9 | 12 |
| 5 | 1 | 0.1452 | 1.186 | 0.062 | 8.09/3 | 0.044 | 8 | 12 |
| 5 | 2 | 0.1497 | 1.051 | 0.218 | 9.20/3 | 0.027 | 8 | 12 |
| 5 | 3 | 0.1555 | 1.168 | 0.025 | 14.69/5 | 0.012 | 10 | 16 |
| 5 | 4 | 0.1575 | 0.935 | 0.032 | 14.56/6 | 0.024 | 9 | 16 |
| 5 | 5 | 0.1585 | 0.853 | 0.032 | 16.93/5 | 0.005 | 4 | 10 |
| 6 | 1 | 0.1460 | 1.138 | 0.071 | 8.63/3 | 0.035 | 8 | 12 |
| 6 | 2 | 0.1505 | 1.005 | 0.048 | 9.84/3 | 0.020 | 8 | 12 |
| 6 | 3 | 0.1562 | 0.925 | 0.029 | 18.35/5 | 0.003 | 4 | 10 |
| 6 | 4 | 0.1583 | 0.856 | 0.033 | 18.97/5 | 0.002 | 4 | 10 |
| 6 | 5 | 0.1593 | 0.824 | 0.036 | 17.19/5 | 0.004 | 4 | 10 |
| 6 | 6 | 0.1600 | 0.806 | 0.045 | 12.93/5 | 0.024 | 4 | 10 |

TABLE X. Best range fits to $M(^1P_1)$ for $am_q = 0.025$ using the cubic group $^1P_1$ state.

| $i$ | $j$ | $(\kappa_i + \kappa_j)/2$ | $aM(^1P_1)$ | error | $\chi^2/N_{DF}$ | conf. lvl. | $t_i$ | $t_f$ |
|---|---|---|---|---|---|---|---|---|
| 1 | 1 | 0.1320 | 1.715 | 0.025 | 0.75/3 | 0.860 | 8 | 12 |
| 2 | 1 | 0.1365 | 1.564 | 0.027 | 1.75/3 | 0.627 | 8 | 12 |
| 2 | 2 | 0.1410 | 1.401 | 0.026 | 1.99/3 | 0.574 | 8 | 12 |
| 3 | 1 | 0.1422 | 1.371 | 0.029 | 3.95/3 | 0.267 | 8 | 12 |
| 3 | 2 | 0.1467 | 1.207 | 0.029 | 4.15/3 | 0.246 | 8 | 12 |
| 3 | 3 | 0.1525 | 0.964 | 0.031 | 2.24/3 | 0.525 | 8 | 12 |
| 4 | 1 | 0.1442 | 1.296 | 0.034 | 4.02/3 | 0.260 | 8 | 12 |
| 4 | 2 | 0.1487 | 1.139 | 0.036 | 3.94/3 | 0.268 | 8 | 12 |
| 4 | 3 | 0.1545 | 0.898 | 0.043 | 2.22/3 | 0.528 | 8 | 12 |
| 4 | 4 | 0.1565 | 0.804 | 0.052 | 1.41/3 | 0.703 | 8 | 12 |
| 5 | 1 | 0.1452 | 1.321 | 0.021 | 12.44/11 | 0.331 | 4 | 16 |
| 5 | 2 | 0.1497 | 1.164 | 0.019 | 11.61/11 | 0.394 | 4 | 16 |
| 5 | 3 | 0.1555 | 0.855 | 0.051 | 7.31/7 | 0.398 | 8 | 16 |
| 5 | 4 | 0.1575 | 0.872 | 0.037 | 12.22/9 | 0.201 | 6 | 16 |
| 5 | 5 | 0.1585 | 0.867 | 0.025 | 15.33/5 | 0.009 | 4 | 10 |
| 6 | 1 | 0.1460 | 1.163 | 0.049 | 5.19/7 | 0.637 | 8 | 16 |
| 6 | 2 | 0.1505 | 1.125 | 0.024 | 9.36/5 | 0.096 | 4 | 10 |
| 6 | 3 | 0.1562 | 0.909 | 0.044 | 3.48/3 | 0.324 | 6 | 10 |
| 6 | 4 | 0.1583 | 0.851 | 0.052 | 3.51/3 | 0.320 | 6 | 10 |
| 6 | 5 | 0.1593 | 0.893 | 0.121 | 3.68/2 | 0.159 | 7 | 10 |
| 6 | 6 | 0.1600 | 0.815 | 0.039 | 13.27/5 | 0.021 | 4 | 10 |

TABLE XI. Best range fits to $M(^1S_0)$ for the quenched data using the pseudoscalar current.

| $i$ | $j$ | $(\kappa_i + \kappa_j)/2$ | $aM(^1S_0)$ | error | $\chi^2/N_{DF}$ | conf. lvl. | $t_i$ | $t_f$ |
|---|---|---|---|---|---|---|---|---|
| 0 | 0 | 0.1300 | 1.411 | 0.002 | 0.88/5 | 0.829 | 17 | 21 |
| 1 | 0 | 0.1375 | 1.119 | 0.002 | 1.52/6 | 0.822 | 17 | 22 |
| 1 | 1 | 0.1450 | 0.804 | 0.002 | 2.06/6 | 0.724 | 17 | 22 |
| 2 | 0 | 0.1410 | 0.978 | 0.003 | 0.58/6 | 0.965 | 17 | 22 |
| 2 | 1 | 0.1485 | 0.650 | 0.002 | 5.34/8 | 0.501 | 15 | 22 |
| 2 | 2 | 0.1520 | 0.479 | 0.002 | 4.30/8 | 0.636 | 10 | 17 |
| 3 | 0 | 0.1420 | 0.937 | 0.003 | 1.36/6 | 0.852 | 17 | 22 |
| 3 | 1 | 0.1495 | 0.607 | 0.002 | 7.73/10 | 0.460 | 12 | 21 |
| 3 | 2 | 0.1530 | 0.421 | 0.002 | 10.57/14 | 0.566 | 9 | 22 |
| 3 | 3 | 0.1540 | 0.364 | 0.003 | 11.31/14 | 0.503 | 9 | 22 |
| 4 | 0 | 0.1425 | 0.917 | 0.004 | 2.55/6 | 0.637 | 17 | 22 |
| 4 | 1 | 0.1500 | 0.586 | 0.002 | 8.23/12 | 0.606 | 10 | 21 |
| 4 | 2 | 0.1535 | 0.393 | 0.002 | 8.61/14 | 0.736 | 9 | 22 |
| 4 | 3 | 0.1545 | 0.331 | 0.002 | 8.96/14 | 0.707 | 9 | 22 |
| 4 | 4 | 0.1550 | 0.298 | 0.003 | 10.68/15 | 0.638 | 8 | 22 |

TABLE XII. Best range fits to $M(^3S_1)$ for the quenched data using the vector current.

| $i$ | $j$ | $(\kappa_i + \kappa_j)/2$ | $aM(^3S_1)$ | error | $\chi^2/N_{DF}$ | conf. lvl. | $t_i$ | $t_f$ |
|---|---|---|---|---|---|---|---|---|
| 0 | 0 | 0.1300 | 1.425 | 0.002 | 3.76/6 | 0.439 | 12 | 17 |
| 1 | 0 | 0.1375 | 1.137 | 0.002 | 7.47/9 | 0.381 | 14 | 22 |
| 1 | 1 | 0.1450 | 0.836 | 0.002 | 6.02/8 | 0.421 | 15 | 22 |
| 2 | 0 | 0.1410 | 1.003 | 0.003 | 6.19/9 | 0.518 | 14 | 22 |
| 2 | 1 | 0.1485 | 0.700 | 0.003 | 4.78/7 | 0.444 | 11 | 17 |
| 2 | 2 | 0.1520 | 0.551 | 0.004 | 4.41/7 | 0.492 | 11 | 17 |
| 3 | 0 | 0.1420 | 0.971 | 0.003 | 6.95/9 | 0.434 | 11 | 19 |
| 3 | 1 | 0.1495 | 0.660 | 0.004 | 2.94/7 | 0.709 | 11 | 17 |
| 3 | 2 | 0.1530 | 0.505 | 0.005 | 3.85/7 | 0.571 | 12 | 18 |
| 3 | 3 | 0.1540 | 0.460 | 0.007 | 4.68/8 | 0.585 | 12 | 19 |
| 4 | 0 | 0.1425 | 0.953 | 0.004 | 3.50/9 | 0.835 | 11 | 19 |
| 4 | 1 | 0.1500 | 0.639 | 0.004 | 3.43/8 | 0.754 | 11 | 18 |
| 4 | 2 | 0.1535 | 0.483 | 0.007 | 4.82/8 | 0.568 | 12 | 19 |
| 4 | 3 | 0.1545 | 0.435 | 0.009 | 4.47/8 | 0.614 | 12 | 19 |
| 4 | 4 | 0.1550 | 0.406 | 0.009 | 7.96/10 | 0.437 | 12 | 21 |

TABLE XIII. Best range fits to $M(^3P_1)$ for the quenched data using the axial vector current.

| $i$ | $j$ | $(\kappa_i + \kappa_j)/2$ | $aM(^3P_1)$ | error | $\chi^2/N_{DF}$ | conf. lvl. | $t_i$ | $t_f$ |
|---|---|---|---|---|---|---|---|---|
| 0 | 0 | 0.1300 | 1.650 | 0.019 | 9.42/13 | 0.583 | 10 | 22 |
| 1 | 0 | 0.1375 | 1.376 | 0.025 | 6.80/13 | 0.815 | 10 | 22 |
| 1 | 1 | 0.1450 | 1.093 | 0.038 | 8.27/13 | 0.689 | 10 | 22 |
| 2 | 0 | 0.1410 | 1.237 | 0.035 | 6.63/13 | 0.828 | 10 | 22 |
| 2 | 1 | 0.1485 | 1.072 | 0.024 | 13.49/16 | 0.489 | 7 | 22 |
| 2 | 2 | 0.1520 | 0.975 | 0.037 | 6.43/10 | 0.599 | 7 | 16 |
| 3 | 0 | 0.1420 | 1.317 | 0.021 | 13.91/16 | 0.456 | 7 | 22 |
| 3 | 1 | 0.1495 | 1.065 | 0.029 | 14.36/14 | 0.278 | 7 | 20 |
| 3 | 2 | 0.1530 | 0.972 | 0.046 | 6.59/10 | 0.582 | 7 | 16 |
| 3 | 3 | 0.1540 | 0.967 | 0.057 | 8.39/10 | 0.396 | 7 | 16 |
| 4 | 0 | 0.1425 | 1.315 | 0.023 | 11.37/16 | 0.657 | 7 | 22 |
| 4 | 1 | 0.1500 | 1.059 | 0.032 | 7.83/9 | 0.348 | 7 | 15 |
| 4 | 2 | 0.1535 | 0.986 | 0.052 | 8.31/10 | 0.404 | 7 | 16 |
| 4 | 3 | 0.1545 | 0.974 | 0.062 | 10.85/10 | 0.210 | 7 | 16 |
| 4 | 4 | 0.1550 | 0.956 | 0.070 | 4.35/6 | 0.360 | 7 | 12 |

TABLE XIV. Mass Splittings and inverse lattice spacings for the $am_q = 0.010$ configurations.

| Mass difference | $\Delta M$ (MeV) | $\kappa$ | $a\Delta M$ | $a^{-1}$ (MeV) |
|---|---|---|---|---|
| Using $\bar{\psi}\gamma_i\gamma_5\psi$ | | | | |
| $a\Delta M(^3P_1 - \overline{S})$ | 442 | 0.1320 | 0.238(9) | 1810(50) |
| | | 0.1410 | 0.248(9) | |
| Using cubic group | | | | |
| $a\Delta M(^3P_1 - \overline{S})$ | 442 | 0.1320 | 0.235(16) | 1800(100) |
| | | 0.1410 | 0.248(20) | |
| $a\Delta M(^1P_1 - \overline{S})$ | 458 | 0.1320 | 0.234(16) | 1900(90) |
| | | 0.1410 | 0.250(18) | |
| $a\Delta M(\overline{P} - \overline{S})$ | 457 | 0.1320 | 0.234(6) | 1900(30) |
| | | 0.1410 | 0.247(7) | |
| Overall estimate of $a^{-1}$ | | | | $1900 \pm 50$ (stat.) $\pm 100$ (sys.) MeV |

TABLE XV. Mass Splittings and inverse lattice spacings for the $am_q = 0.025$ configurations.

| Mass difference | $\Delta M$ (MeV) | $\kappa$ | $a\Delta M$ | $a^{-1}$ (MeV) |
|---|---|---|---|---|
| Using $\bar{\psi}\gamma_i\gamma_5\psi$ | | | | |
| $a\Delta M(^3P_1 - \overline{S})$ | 442 | 0.1320 | 0.257(10) | 1660(40) |
| | | 0.1410 | 0.274(11) | |
| Using cubic group | | | | |
| $a\Delta M(^3P_1 - \overline{S})$ | 442 | 0.1320 | 0.256(21) | 1580(120) |
| | | 0.1410 | 0.302(36) | |
| $a\Delta M(^1P_1 - \overline{S})$ | 458 | 0.1320 | 0.249(17) | 1720(100) |
| | | 0.1410 | 0.284(25) | |
| $a\Delta M(\overline{P} - \overline{S})$ | 457 | 0.1320 | 0.258(26) | 1660(110) |
| | | 0.1410 | 0.293(26) | |
| Overall estimate of $a^{-1}$ | | | | $1660 \pm 110$ (stat.) $\pm 100$ (sys.) MeV |

TABLE XVI. Inverse lattice spacings (in MeV) using other quantities.

| $am_q$ | "force" | $m_\rho$(W) | $m_p$(W) | $m_\rho$(S) | $m_P$(S) | NRQCD |
|---|---|---|---|---|---|---|
| 0.025 | 1935 | 2000 | 1685 | — | — | — |
| 0.010 | 2055 | 2140 | 1800 | — | — | 2400 |
| 0.0 | 2135 | 2230 | 1875 | 1800 | 1660 | — |

TABLE XVII. Mass Splittings and inverse lattice spacings for the quenched ($am_q = \infty$) configurations.

| Mass difference | $\Delta M$ (MeV) | $\kappa$ | $a\Delta M$ | $a^{-1}$ (MeV) |
|---|---|---|---|---|
| $a\Delta M(^3P_1 - \overline{S})$ | 442 | 0.1300 | 0.23(3) | 1800(150) |
| | | 0.1450 | 0.26(3) | |
| Overall estimate of $a^{-1}$ | | | | $1800 \pm 150$ (stat.) $\pm 100$ (sys.) MeV |

TABLE XVIII. Lattice and experimental determinations of the masses of the charmed heavy-light mesons ($am_q = 0.010$).

| meson | lattice (MeV) | expt. (MeV) |
|---|---|---|
| $D^0$ | 1700(100) | 1864 |
| $D^*$ | 1750(100) | 2010 |
| $D_1$ | 2250(130) | 2423 |

TABLE XIX. Lattice and experimental determinations of the masses of the charmed heavy-light mesons ($am_q = 0.025$).

| meson | lattice (MeV) | expt. (MeV) |
|---|---|---|
| $D^0$ | 1690(110) | 1864 |
| $D^*$ | 1730(110) | 2010 |
| $D_1$ | 2090(140) | 2423 |

TABLE XX. Lattice and experimental determinations of the masses of the charmed heavy-light mesons ($am_q = \infty$).

| meson | lattice (MeV) | expt. (MeV) |
|---|---|---|
| $D^0$ | 1670(140) | 1864 |
| $D^*$ | 1740(140) | 2010 |
| $D_1$ | 2050(170) | 2423 |

TABLE XXI. Lattice and experimental determinations of the mass differences of the charmed heavy-light mesons ($am_q = 0.010$).

| mass difference | lattice (MeV) | expt. (MeV) |
|---|---|---|
| $M(D^*) - M(D^0)$ | 50(3) | 146 |
| $M(D_1) - [M(D^0) + 3M(D^*)]/4$ | 560(30) | 450 |

TABLE XXII. Lattice and experimental determinations of the mass differences of the charmed heavy-light mesons ($am_q = \infty$).

| mass difference | lattice (MeV) | expt. (MeV) |
|---|---|---|
| $M(D^*) - M(D^0)$ | 38(3) | 146 |
| $M(D_1) - [M(D^0) + 3M(D^*)]/4$ | 670(60) | 450 |